\documentclass[nohyper,article,11pt,notoc]{JHEP3}

\pdfoutput=1
\usepackage{graphicx}
\usepackage{amssymb}
\usepackage{amsmath}
\usepackage{amsfonts}
\usepackage{latexsym}
\usepackage[all]{xy}
\usepackage{slashed}
\usepackage{slashbox}
\usepackage[numbers ,square, comma, sort&compress]{natbib} 
\usepackage{xfrac}

\preprint{}

\def\ra{\rightarrow}

\newcommand{\PZz}{\mathrm{Z}}

\newcommand{\GEV}{\mbox{$\mathrm{~GeV}$}}

\newcommand{\GEVcc}{\mbox{$\mathrm{~GeV}/{{\it c}^2}$}}

\newcommand{\bbbar}{\mbox{$\mathrm{b\bar{b}}$}}

\newcommand{\ma}{m_\A}
\newcommand{\mh}{m_\mathrm{h}}
\newcommand{\mA}{m_\mathrm{a}}

\newcommand{\A}{\mathrm{a}}

\title{\center{RECAST \\ \large{Extending the Impact of Existing Analyses
}}}
\author{Kyle Cranmer and Itay  Yavin\\ \it{Center for Cosmology and Particle Physics, Department of Physics, New York University, New York, NY 10003}\\} 

\abstract{
Searches for new physics by experimental collaborations represent a significant investment in time and resources.  Often these searches are sensitive to a broader class of models than they were originally designed to test.  
We aim to extend the impact of existing searches through a technique we call \textit{recasting}. 
After considering several examples, which illustrate the issues and subtleties involved, we present RECAST, a framework designed to facilitate the usage of this technique.
\\
\\
\textsc{Website: } \textit{www.recast.it}
}

\begin{document}

\section{Introduction}

Over the past several decades, many extensions and alterations to the  standard model (SM) of particle physics have been proposed. These take the form of concrete modifications or extensions to the known particle spectrum and interactions; resulting in predictions that are testable at the energy and intensity frontiers found in high energy colliders.  Particle experimentalists design and conduct searches for new physics at these colliders, which requires a significant investment of time and resources. Often these searches are sensitive to a broader class of models than they were originally designed to test, thus it is natural to ask
\\

\textit{What impact does an existing analysis have on an alternative signal hypothesis?} 
\\

The ability to accurately answer this question would extend the impact of our existing searches with little additional effort.  If one sacrifices the optimality of a dedicated search, then one can reuse the estimates of backgrounds and systematic uncertainties from the original search as well as the observations in data.  The only piece of information necessary to \textit{recast} the results of an existing analysis into the context of a new theory is the expected signal yield for that model.  We call this technique \textit{recasting}, and in Sec.~\ref{sec:examples} we consider several examples where this has been successfully done.  The advantages of this approach are that it
\begin{list}{\labelitemi}{\leftmargin=1.5em}
 \item extends the impact of existing results from experimental collaborations,
 \item provides accurate interpretation of existing searches in the context of alternative models,
 \item does not require access to or reprocessing of the data,
 \item does not involve design of new event selection criteria, and
 \item does not require additional estimates of background rates or systematic uncertainties.
\end{list}

Despite the simplicity of the question above, and its \textsl{in principle} straightforward resolution, it is presently difficult to answer \textsl{in practice}. This difficulty is not the result of a lack of tools, which are to a large extent available, but rather it is because a framework where such a question can be easily asked and accurately answered is not currently available. In Sec.~\ref{S:implementation} we present RECAST, a framework designed to facilitate this type of analysis.  The framework has been designed so that it will
\begin{list}{\labelitemi}{\leftmargin=1.5em}
 \item connect those interested in alternative signals with those responsible for relevant searches,
 \item standardize the format of such requests,
 \item maintain collaborations' control over the approval of new results,
  \item allow new models to be considered even after a search is completed, and
 \item complement data archival efforts.
\end{list}

For completeness, let us briefly review the typical steps involved in the comparison of a new theory with experiment. First, one must simulate the hard scattering processes associated with the new physics phenomena.  The hard scattering process must be supplemented by the parton distribution functions, initial and final state radiation, the process of hadronization, and etc.~\cite{Caveat}.  Next, the interaction of the resulting collection of physical particles with the detector must be simulated, typically using GEANT.  Then the same algorithms used to analyze real data are used to reconstruct the simulated events.  Finally, one is ready to determine the expected yield of signal events satisfying an analysis's event selection criteria, which are considered to be fixed a priori for the purposes of this paper.

Important physical effects may enter at each of these stages, which is why the field has devoted so much effort to the requisite tools.  Fortunately, the detector simulation and reconstruction algorithms for a given experiment are general purpose, so that preparing simulated samples for different processes is largely automated.  Furthermore, standard interfaces have been developed so that simulated events at the parton-level can be developed to the hadron-level and so that simulated events at the hadron-level can be fed to the detector simulation and reconstruction algorithms.  Tools such as MadGraph~\cite{Maltoni:2002qb} are now capable of generating these parton-level samples for fairly general processes simply by specifying particle content and interactions in a suitably general class of theories.

The final test of a new theory brings together the observed yield of events in data, the expected yield from signal processes, the expected yield from background processes, and the uncertainties on these estimates.  Estimating the background and its uncertainty is typically more involved than the process described above for estimating the signal yield.  This is because the event selection criteria select extreme tails of the background processes.  When backgrounds are estimated from theoretical predictions, it often requires more sophisticated modeling than the signal processes.   Other background processes are estimated using data-driven techniques, which are intertwined with the event selection criteria itself.  The design of the event selection criteria is effectively an art that aims at optimizing the sensitivity for some new physics signal in light of these considerations.  Given the hard work that went into this process, it is natural to reuse and extend the impact of existing searches.

\section{Motivation}

We devote this section to try and convince the reader that the ability to recast past searches may benefit all members of the particle physics community who are directly involved with the search for new physics, both experimentalists and theorists. In what follows, we divide our considerations into two separate periods in the lifetime of an experiment, namely \textsl{before} and \textsl{after} discovery. We do so in order to illustrate the extended service provided by RECAST as well as the framework's continuous utility throughout the different phases of the scientific endeavor and search for new physics. 

We emphasize that RECAST is intended for reusing and recasting \textsl{existing} analyses. It is not aimed at, nor even capable of, participating in the discovery process itself. Discovery, undoubtedly the most exciting part of any exploration, must be achieved through the efforts and ingenuity of the experimentalists themselves.

\begin{figure}[h]
\begin{center}
\vspace{-.1in}
\includegraphics[scale=.75]{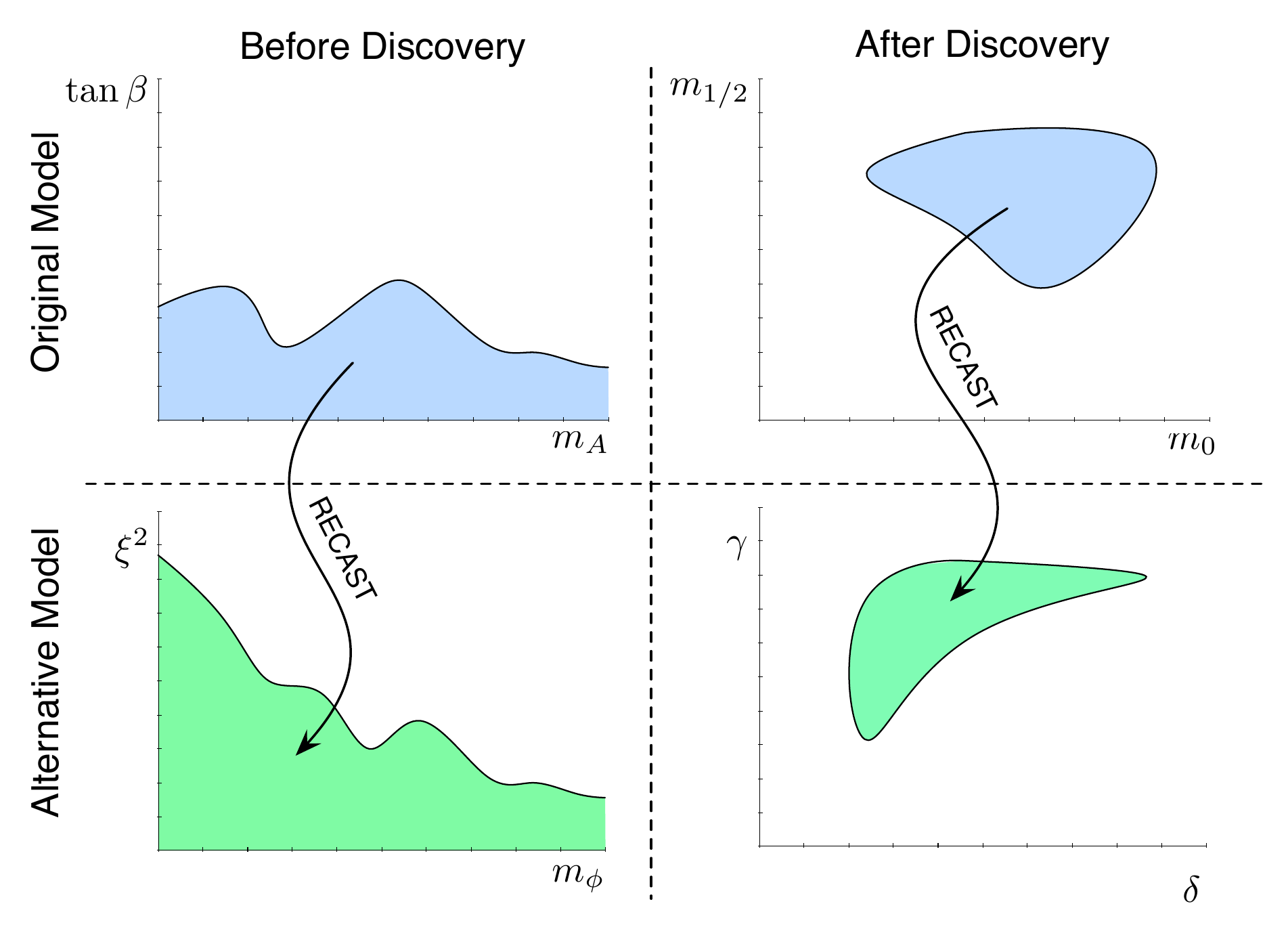}
\end{center}
\caption{A schematic diagram showing the different usages of RECAST. \textit{Before} the discovery of new physics it can be used to translate exclusion regions between different parameter spaces. \textit{After} the discovery of new physics it can be employed to examine the viability of an alternative explanation for the observed distributions.}
\label{fig:schematic}
\end{figure}

\subsection*{Before Discovery}

First, let us consider Eve, an Experimentalist who designs and performs a search for a particular signal. She may be motivated by a particular manifestation of supersymmetry or large extra dimensions, and chooses her cuts accordingly. Supposing she found no excess of events over the SM background processes, she can exclude those signal hypotheses with a certain degree of confidence. By incorporating her analysis into RECAST, she can broaden the impact of her search beyond the particular signal she originally intended. If her collaboration provides an archival system that interfaces with RECAST, then she leaves open the possibility to impact future models that may arise.

Oscar, the Other experimentalist we consider, may be interested in a search for a different signal. Oscar may first wish to ascertain that the signal he is interested in is not already excluded by Eve's search. If Eve integrates her search into RECAST, then he can use it to study his signal. Parts of the parameter space of the signal Oscar is interested in may already be covered by Eve's search (high efficiency) whereas other parts may not (low efficiency). Oscar can now tune his analysis to target those parts of the parameter space which are still viable.  

Finally, consider a Theorist named Theodora, who may have an alternative signal in mind. This particular signal has not been directly searched for, but Eve's search is possibly relevant. She can approach the experimental community and gather interest in this new scenario by initiating a quantitative dialogue through RECAST. 

Eve, Oscar, and Theodora all benefit from the ability to RECAST the original exclusion limits of an existing search to exclusion limits for an alternative hypothesis. These different persona represent, in broad brush strokes, some of the possible ways particle physicists may engage with experimental searches for new physics in those cases where an exclusion limit is reported (which naturally constitute the majority of searches). 

\subsection*{After Discovery}

Suppose now that Eve, the experimentalist, establishes a significant excess above SM background in a particular search she conducts. The collaboration claims discovery of new physics and shows that the data is indeed consistent with a certain region of the minimal supersymmetric SM (MSSM) parameter space.  At this stage, other members of the community would likely be interested to know whether alternative signals are also consistent with the data.  If Theodora can demonstrate that her model for new physics is a viable alternative explanation, then that will garner interest in that scenario.  It may even motivate Oscar, the other experimentalist, to design an analysis to differentiate between the competing explanations.

The process of constructing and eliminating competing hypotheses may continue for a while before a new and established paradigm emerges. Using RECAST we can more efficiently obtain answers to questions that we naturally ask during this process. It allows for a more complete consideration and better integration of \textsl{existing} analyses and their impact on the different competing proposals.

\section{Examples and Case-Studies}\label{sec:examples}

In this section we review several case-studies that exemplify and illustrate the power and utility of recasting existing analyses. Our examples include past published works such as the LEP compendium of Higgs boson searches as well as our own recasting of an experimental search we recently conducted.  

\subsection{Compendium of neutral Higgs boson searches at LEP}
\label{sec:LEPCompendium}
The different collaborations at LEP performed extensive searches for the Higgs boson and excluded SM scenarios  with masses below $114.4~\GEVcc~$\cite{Barate:2003sz}.  Most of the searches were based on Higgs decaying into $\bbbar$ or $\tau^+\tau^-$; however, final states specific to the MSSM and more generic scenarios were also considered.  The LEP working group for Higgs boson searches produced a compendium of results based on these searches, presented as limits on the production cross-section times branching ratio for specific production and decay modes as a function of the physical Higgs masses~\cite{Schael:2006cr}.  The results were then interpreted in several MSSM benchmarks, including CP-violating ones, and integrated into \texttt{HiggsBounds}~\cite{Bechtle:2008jh},  a program used to check the consistency of new Higgs sectors with the LEP constraints.

In the process of preparing the compendium, many of the previous searches were extended to different production and decay modes or additional Higgs mass scenarios.  In many cases this did not involve designing a new search, but simply evaluating the efficiency of existing searches to alternative signal scenarios.  For instance, in Ref.~\cite{Abbiendi:2004ww} the OPAL collaboration used the $e^+e^- \to HZ \to b\bar{b}\,q\bar{q}$ analysis with identical event selection criteria to target the alternative signal involving the cascade $\mathcal{H}_2\to\mathcal{H}_1\mathcal{H}_1$.  Table 3 of Ref.~\cite{Abbiendi:2004ww}  shows the efficiencies of the standard search for $e^+e^- \to HZ \to b\bar{b}\,q\bar{q}$ to the alternative $e^+e^- \to \mathcal{H}_2 Z\to  \mathcal{H}_1\mathcal{H}_1 Z \to  b\bar{b}\,b\bar{b}\,q\bar{q}$ signal.   The existing analysis was highly efficient for this alternative process, supporting our claim that existing analyses are often sensitive to a broader class of models than they were initially intended.  Similar studies were carried out by DELPHI~\cite{Delphi:2007ge} to determine efficiency of the LEP1 search for $\tau^+\tau^-\,b\bar{b}$ to the $hA\to\tau^+\tau^- q\bar{q}$ signal and the efficiency of a search for $h\to q\bar{q}$ to the $hZ\to AAZ \to c\bar{c}\, c\bar{c}\, q\bar{q}$ signal.
%
%

These examples  demonstrate the ability to extend the impact of existing analyses; the sensitivity of existing analyses to alternative signals with different final state topologies; the ability to accurately calculate efficiencies even when complicated algorithms such as neural networks are involved; and the acceptance of such results by the community when they are provided directly by the collaborations.  The examples also document how such an approach streamlines the work of experimentalists, as new analyses were not designed when existing analyses were sufficiently sensitive.

It should be emphasized that estimating the efficiency of the existing analyses for alternative signals was only possible because the codes used for the original analyses as well as the necessary detector simulation and reconstruction algorithms were still available and functioning.  At this point in time it would be difficult, if not impossible, to assess the sensitivity of these analyses to a new alternative signal.  As a result, several exotic decays, such as those considered in Refs.~\cite{Dermisek:2005ar,Dermisek:2007yt,Chang:2005ht,Chang:2008cw,Bellazzini:2009xt,Bellazzini:2009kw} are not constrained in a precise way.  


The RECAST framework aims to recreate and build upon the success of these examples by providing a standard interface for such queries and by providing an archival system that can execute all necessary analysis codes to support future requests.  We expand on the design of the framework further in Sec.~\ref{S:implementation}, but point out here that the implementation of the back-end that actually archives the analysis codes is the responsibility of the collaboration.  

\subsection{Searches for exotic decays of a light Higgs boson at ALEPH}

The mild tension with electroweak precision tests~\cite{Barbieri:1999tm}, the small excess in the $\bbbar$ final state observed at LEP~\cite{Dermisek:2005gg}, and the fine-tuning needed in the MSSM have prompted the consideration of models with exotic Higgs boson decays~\cite{Dermisek:2005ar,Dermisek:2007yt,Chang:2005ht,Chang:2008cw,Bellazzini:2009xt,Bellazzini:2009kw}. In these models, new decay channels can dominate over $\mathrm{h} \ra \bbbar$ and render the Higgs boson  ``invisible'' for conventional searches. The decay of the Higgs boson into a pair of light pseudoscalars is particularly well motivated by these constructions. 


Over the past year the authors of the present paper were involved in a search for a light Higgs boson (with a mass $\mh < 114\GEVcc$) decaying into a four-tau final state using data collected by the ALEPH detector during the LEP2 run~\cite{Schael:2010aw}. The search directly targeted the process $e^+e^-\ra \mathrm{Z}\mathrm{h}$ with $\mathrm{h}\ra \A\A \ra 4\tau$'s and $\mathrm{Z}\ra e^+e^-,~\mu^+\mu^-, \bar{\nu}\nu$. No excess of events above background was seen and exclusion limits on the production cross-section and branching ratio into four taus were reported. We use this search as a case-study to illustrate the different facets of RECAST\footnote{It is possible and even likely that some of the past searches~\cite{Barate:2003sz,Schael:2006cr} are sufficiently sensitive to the different final states we discuss in this section to render some of these possibilities excluded already. But, since no reported search have looked at these exact decay modes, and no system has archived the analysis, it is impossible to accurately assess how excluded these possibilities really are. It is precisely for this purpose that RECAST  has been developed.}.

\subsubsection*{Other leptonic decays of the pseudo-scalar}

The original analysis treated the decay products of $h\to aa$ as two jets and required that each jet had two or four tracks\footnote{Since the pseudo-scalars are light, $\ma \lesssim 10\GEVcc$, they are fairly boosted and their decay products are collimated. Each $\tau$ decays into "one-prong" (85\%) or "three-prongs" (15\%) which results in 2, 4, or 6 nearby tracks for each pseudo-scalar.}. Thus it is natural to ask how sensitive this analysis is to the decay of the pseudo-scalar into  electrons or muons.  In Table \ref{tbl:ExtendedImpact} we present the efficiency for the original decay considered as well as the alternative lepton decays of the pseudo-scalar. Indeed, as expected, the search is fairly sensitive to these alternative decay modes as well. The $\A\ra \mu^+\mu^-$ is in fact as efficient  as the original signal. The impact of the search is thus extended to cover alternative signals. 
Table \ref{tbl:ExtendedImpact} also shows new results from recasting the original analysis into these final states expressed as  95\% upper-limits on  $\xi^2 = \sfrac{\sigma(e^+ e^-\ra Z+h)}{\sigma_{SM}(e^+ e^-\ra Z+h)}\times\mathrm{B}(h\ra aa)\times\mathrm{B}(a\ra l^+l^-)^2 $.


The $\A\ra e^+e^-$ case serves as a reminder that subtle effects may enter into the determination of the final efficiency. It demonstrates the necessity for an accurate detector simulation and faithful reprocessing of the alternative signal through \textsl{all} the original cuts. One might naively expect the electron decay channel to have the same efficiency as the muon channel, but in fact it is lower by about 30\%. This is \textsl{not} due to electron vs. muon detection efficiencies, but rather an indirect consequence of enhanced brehmstrahlung from electrons due to the magnetic field in the detector. The radiated photons are often not included in the two jets.  As a result the events often fail the requirement that $E_{\rm j_1}+ E_{\rm j_2}+\slashed{E} > E_{\rm CM} - 5~\GEV$, where $ E_{\rm j_{1,2}}$ is the energy in the two jets formed from the $\A$'s decay and $ \slashed{E} $ is the missing energy in the event. 
  
This example reinforces the need for a framework like RECAST if we  wish to accurately obtain the efficiency associated with alternative signals. It illustrates the danger in estimating the efficiency of other signals through anything but a realistic detector simulation and reconstruction of the events and a complete account of \textsl{all} the cuts employed in the original analysis. 
 
\begin{table}[h]
\caption{New results from recasting the ALEPH analysis of Ref.~\cite{Schael:2010aw} to leptonic decay modes for the pseudo-scalar with $\mh=100\GEVcc$, $\mA=10 \GEVcc $. The total signal efficiency as well as 95\% confidence level limits on $\xi^2 = \sfrac{\sigma(e^+ e^-\ra Z+h)}{\sigma_{SM}(e^+ e^-\ra Z+h)}\times\mathrm{B}(h\ra aa)\times\mathrm{B}(a\ra l^+l^-)^2 $ are shown. We note the lower efficiency in the case of $\A\ra e^+e^-$, which is explained in the text.}
\begin{center}
ALEPH Archived Data\\
\vspace{2mm}
\begin{tabular}{|c|c|c|}
\hline
Decay mode & Efficiency & $\xi^2$ \\
\hline
$\A\ra\tau^+\tau^-$  &  0.37 & 0.46\\ \hline
$\A\ra \mu^+\mu^-$ & 0.35 & 0.14\\
$\A\ra e^+e^-$ & 0.27 & 0.20\\
\hline
\end{tabular}
\end{center}
\label{tbl:ExtendedImpact}
\end{table}%

\subsubsection*{Mixed decays of the pseudo-scalars} 

The limits on $\mathrm{h}\ra\A\A\ra 4\tau$'s leave open the possibility that the pseudo-scalar decays a substantial fraction of the time into other light SM particles, e.g. gluons or charm quarks. In this case, the mixed decay, namely $\mathrm{h}\ra\A\A\ra \tau^+\tau^-~gg$, or $\mathrm{h}\ra\A\A\ra \tau^+\tau^-~c\bar{c}$, might actually dominate over any of the pure channels where both pseudo-scalar decay in the same fashion. To completely close the window on the scenario of Ref.~\cite{Dermisek:2005gg} and exclude the entire parameter space, it is necessary to rule-out the mixed decay as well. 

\begin{figure}[t]
\begin{center}
\vspace{-.1in}
\includegraphics[scale=.85]{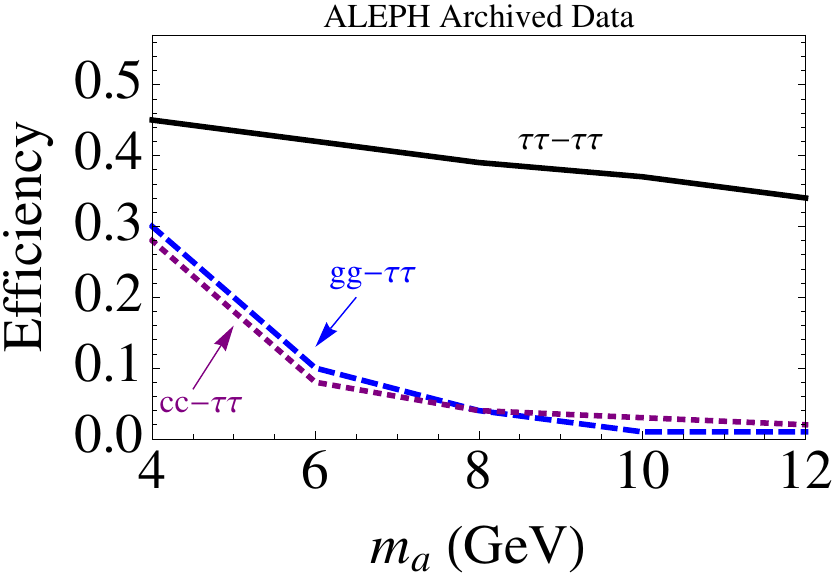}
\includegraphics[scale=.85]{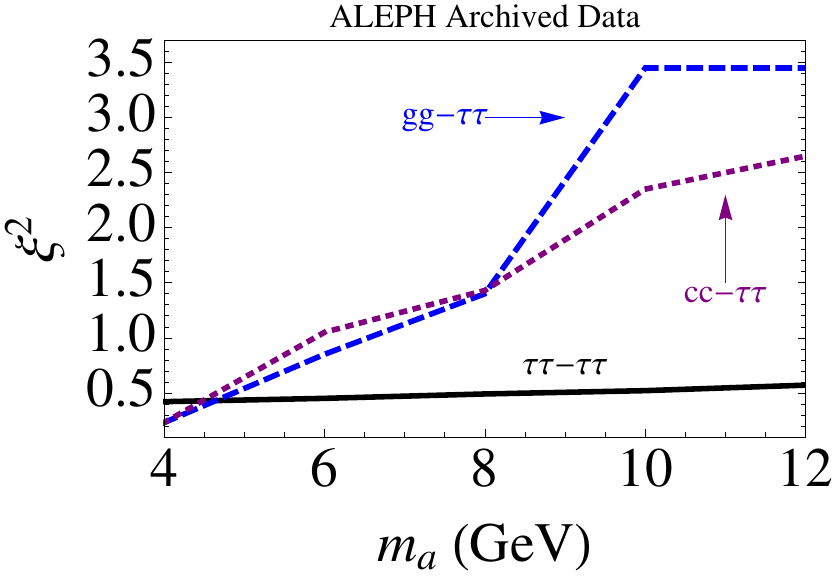}
\end{center}
\caption{New results from recasting the ALEPH analysis of Ref.~\cite{Schael:2010aw} to mixed decay modes for the pseudo-scalar. On the left pane we plot a comparison of the efficiency of the search of Ref.~\cite{Schael:2010aw} for $\mathrm{h}\ra\A\A\ra 4\tau$'s to the alternative mixed decay modes $\mathrm{h}\ra\A\A\ra 2g\tau^+\tau^-$ (blue-dashed), and $\mathrm{h}\ra\A\A\ra c\bar{c}\tau^+\tau^- $ (purple-dotted) with $\mh=100\GEVcc$ for the $\PZz\rightarrow \nu\bar{\nu}$. The efficiency for the channels with leptonic decays of the $\PZz$ are effectively zero. On the right pane we plot the 95\% CL limit on $\xi^2 = \sfrac{\sigma(e^+ e^-\ra Z+h)}{\sigma_{SM}}\times\mathrm{B}(h\ra aa)\times\mathrm{B}(a\ra \tau^+\tau^-)\times\mathrm{B}(a\ra gg~\mathrm{or}~ c\bar{c})$. The decrease in efficiency for the mixed decay with higher pseudo-scalar mass is described in the text. }
\label{fig:MixedChannelEfficiency}
\end{figure}

Just like Oscar, in our hypothetical story above, we wanted to know how sensitive the original $4\tau$ search would be to the mixed channel.  We recast the original analysis for these mixed decays, and in Fig. \ref{fig:MixedChannelEfficiency} we present the efficiency of the analysis to these alternative signals in comparison with the original signal as a function of the pseudo-scalar mass. The most striking feature to note is the precipitous drop in efficiency for the alternative scenarios with heavier pseudo-scalar masses. This loss of efficiency is due to the requirement that each jet have two or four tracks. For the alternative signals, where one of the $\A$'s decays hadronically, the number of tracks in the corresponding jet rises with the mass of this pseudo-scalar. Here the important effects are in the parton shower and track reconstruction.


By recasting the analysis, we have established that the low $\mA$ region of the parameter space is well-constrained by the original $\mathrm{h}\ra\A\A\ra 4\tau$'s analysis. When designing the search for the mixed decay we are likely to place more emphasis on the higher $\mA$ region where a signal could really be present rather than waste effort on a part of the parameter space already constrained by a previous search. 

\subsubsection*{Hadronic decays of the pseudo-scalars}
 
Our last example related to the $\mathrm{h}\ra\A\A\ra4\tau$'s search illustrates why an archival framework like RECAST would be beneficial, namely the ability to assess the viability of new theoretical developments in light of existing analyses. 

Following the original proposal of Ref.~\cite{Dermisek:2005ar} that a Higgs boson decaying into 4$\tau$'s may have been missed by the original LEP searches, Chang et al.~\cite{Chang:2005ht} and later Bellazzini et al.~\cite{Bellazzini:2009xt,Bellazzini:2009kw} proposed the decay into 4 gluons or 4 charms as other Higgs decays which might have been missed. The flavor independent searches for $h\to q\bar{q}$  at LEP~\cite{Heister:2002cg,Abdallah:2004bb,Abbiendi:2004gn,Achard:2003ty,Gaycken:2003gz} are likely to have been sensitive to these alternative signals, but the peculiarities of the alternative signals and the absence of any framework like RECAST make it impossible to quantify how viable such scenarios really are. 

\begin{figure}[t]
\begin{center}
\vspace{.1in}
\includegraphics[scale=.85]{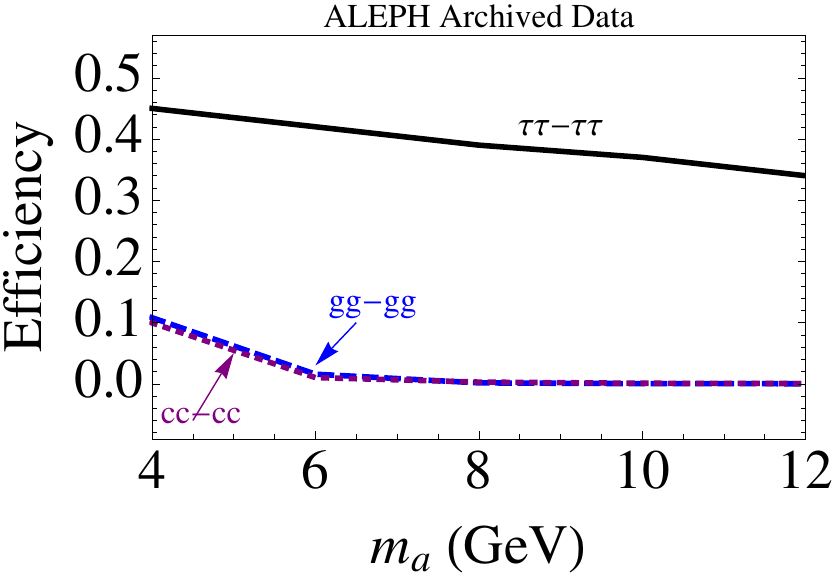}
\includegraphics[scale=.85]{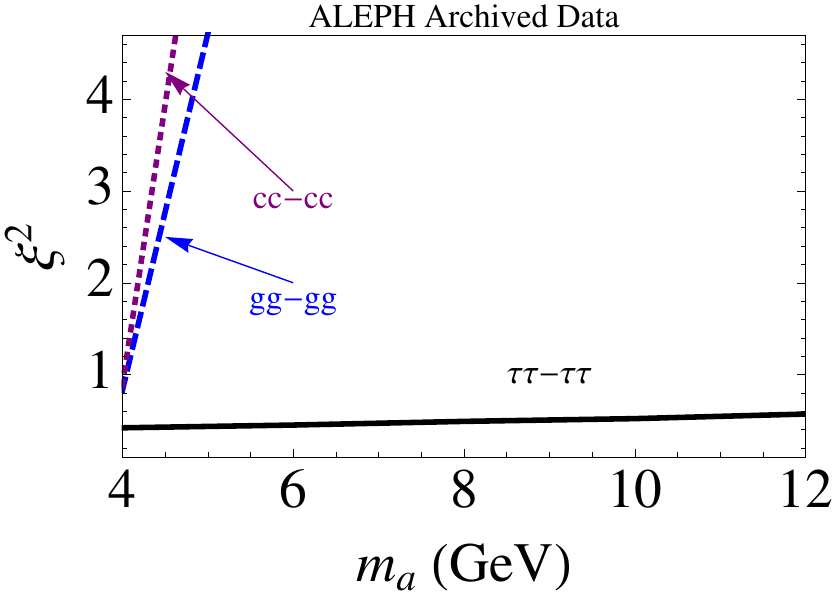}
\end{center}
\caption{New results from recasting the ALEPH analysis of Ref.~\cite{Schael:2010aw} to fully hadronic decay modes for the pseudo-scalar.  A comparison of the efficiency of the search of Ref.~\cite{Schael:2010aw} for $\mathrm{h}\ra\A\A\ra 4\tau$'s to the alternative fully hadronic decay modes $\mathrm{h}\ra\A\A\ra 4g$ (blue-dashed), and $\mathrm{h}\ra\A\A\ra 4c $ (purple-dotted) with $\mh=100\GEVcc$ for the $\PZz\rightarrow \nu\bar{\nu}$. The efficiency for the channels with leptonic decays of the $\PZz$ are effectively zero.. . On the right pane we plot the 95\% CL limit on $\xi^2 = \sfrac{\sigma(e^+ e^-\ra Z+h)}{\sigma_{SM}}\times\mathrm{B}(h\ra aa)\times\mathrm{B}(a\ra gg~\mathrm{or}~ c\bar{c})^2$.  The much lower efficiency exhibited by the fully hadronic channels is explained in the text.}
\label{fig:FullyHadronicEfficiency}
\end{figure}

We also recast the $4\tau$ search~\cite{Schael:2010aw} for these fully hadronic final states and found that it is not sensitive for such decays.  Figure~\ref{fig:FullyHadronicEfficiency} shows the efficiency of the fully hadronic decays. The low efficiency is largely due to two effects. First, , as in the case of the mixed decays, the charge multiplicity for the hadronic decays rises with the pseudo-scalar mass and fails the track multiplicity requirement. Second, since the only major source of missing mass in this case is the $\rm{Z}$ itself, the missing mass distribution is peaked at around $91\GEVcc$ and is fairly narrow. Therefore, many of the alternative signal events fail the rather stringent cut on missing mass ($\slashed{m}> 90\GEVcc$) present in the original 4$\tau$ analysis. 

\newpage
\subsection{Searches for General Gauge Mediation in the Tevatron}

Recent progress~\cite{Cheung:2007es,Meade:2008wd} in gauge-mediated supersymmetry breaking (GMSB) model building has made it clear that the low-energy phenomenology of these models can differ markedly from the well-studied scenario of ordinary GMSB~\cite{Dine:1981za,Dimopoulos:1981au}. Motivated by these developments, the authors of Ref.~\cite{Meade:2009qv} have carried out an impressive survey of existing Tevatron searches sensitive to the extended phenomenology exhibited by general GMSB~\cite{Meade:2008wd}.  

The authors of~\cite{Meade:2009qv} first discuss the classical search mode for ordinary GMSB, namely searches for $\gamma\gamma\slashed{E}_T$ and their relevance to the more general scenarios. The authors based their results on the signal efficiency that was quoted by the original experimental search~\cite{Aaltonen:2009tp}.   However, when the authors used  PGS~\cite{pgs}, a generic detector simulator, and a recreation of the analysis in Ref.~\cite{Aaltonen:2009tp} to evaluate these scenarios, they noted that the efficiency could change by as much as a factor of two as they scan over the parameter space due to photon isolation requirements.  In their case  the cross-section and branching ratio varied so rapidly across the parameter space that the results were robust against this uncertainty in the signal efficiency.  If their result was more sensitive to the variation signal efficiency, then one would not want to rely on the estimate of a subtle effect like photon isolation from anything but the collaboration's detector simulation.  

Ref.~\cite{Meade:2009qv} also investigated in detail other promising channels such as $\gamma+W+\slashed{E}_T$, $Z(l^+l^-)\slashed{E}_T+X$, and etc. In these cases, the authors had to rely on the signal efficiency on their own using the generic detector simulator PGS~\cite{pgs}. They found PGS to be reasonably accurate as a simulator, except for processes that involved instrumental effects like fake leptons or fake missing ${E}_T$. 
Once again, if those searches were incorporated into a framework like RECAST, then the authors could have avoided the labor-intensive and error-prone undertaking of calculating the efficiency of a given search to an alternative signal, and the recast results would have carried the full authority of the collaborations.
%

RECAST is \textit{not} designed to, nor even capable of helping with the other issues addressed by Ref.~\cite{Meade:2009qv}, namely: the identification of alternative search modes; the optimization of existing analyses to target these modes; and the relaxation of existing cuts to allow for more general signatures.  Instead, RECAST allows researches to concentrate on these creative aspects knowing that they have properly accounted for the impact of existing searches.


\subsection{Other examples}

There are several other examples of strategies that have been proposed for coping with the plethora of new physics scenarios of interest to the community, here we mention just a few. In Ref.~\cite{Dube:2008kf} the authors  present a technique for interpreting the results of tri-lepton searches at the Tevatron in a model-independent way. It bares similarity to \texttt{HiggsBounds} (see Sec.~\ref{sec:LEPCompendium}), and is similarly constrained to alternative signals with the same final state. The D$\slashed{0}$ collaborations presented Quaero, an automated system for confronting alternative signals with collider data~\cite{Abazov:2001ny}.  In contrast to RECAST, Quaero automates the production new, optimized analyses; however, this opens the door to new systematic effects, new backgrounds, and other subtleties.  A more restricted version of this algorithm was presented in Ref.~\cite{Caron:2006fg}.  More recently, the authors of Ref.~\cite{Flacco:2010rg} generalized the limits on fourth generation quarks coming from CDF to various two-flavor scenarios under certain assumptions regarding the signal efficiency.  RECAST has shared motivation with these examples and aims to provide an authoritative result under the aegis of the experimental collaborations.


\section{Design and Implementation}\label{S:implementation}

In this section we propose a high-level design for the RECAST framework and discuss some issues that must be addressed in its implementation.  Immediately, we factorize the system into three main components.
\begin{itemize}
\item \textbf{The front-end} which serves as a communication broker: provides a web interface for collecting requests, performs a basic match-making role to connect requests and analyses, and manages inputs and approved results.  The front-end has no authority and no direct access to the analyses.
\item \textbf{The back-ends} which serve as the workhorses of the system: processes an alternative signal through an archived analysis chain, determines signal efficiencies and limits on production rate, provides authority of the result.  Several back-ends are anticipated, the implementation of each being the responsibility of a particular collaboration.
\item \textbf{The API} which defines the interface between the front-end and the back-end.  A well-defined API (application programming interface) allows for multiple back-end implementations to work seamlessly with a single front-end.  It also allows for the back-end to evolve from a manual system to a fully automated system without affecting the front-end.
\end{itemize}
We stress that the framework \textit{does not} need or have access to the data, \textit{does not} involve design of new analyses, and \textit{does not} require additional estimates of background rates or systematics.  We also stress that the authority of any new results is derived from the collaborations themselves and that the original analyses should be the primary citation.

\begin{figure}[h]
\begin{center}
\vspace{-.1in}
\includegraphics[width=\textwidth]{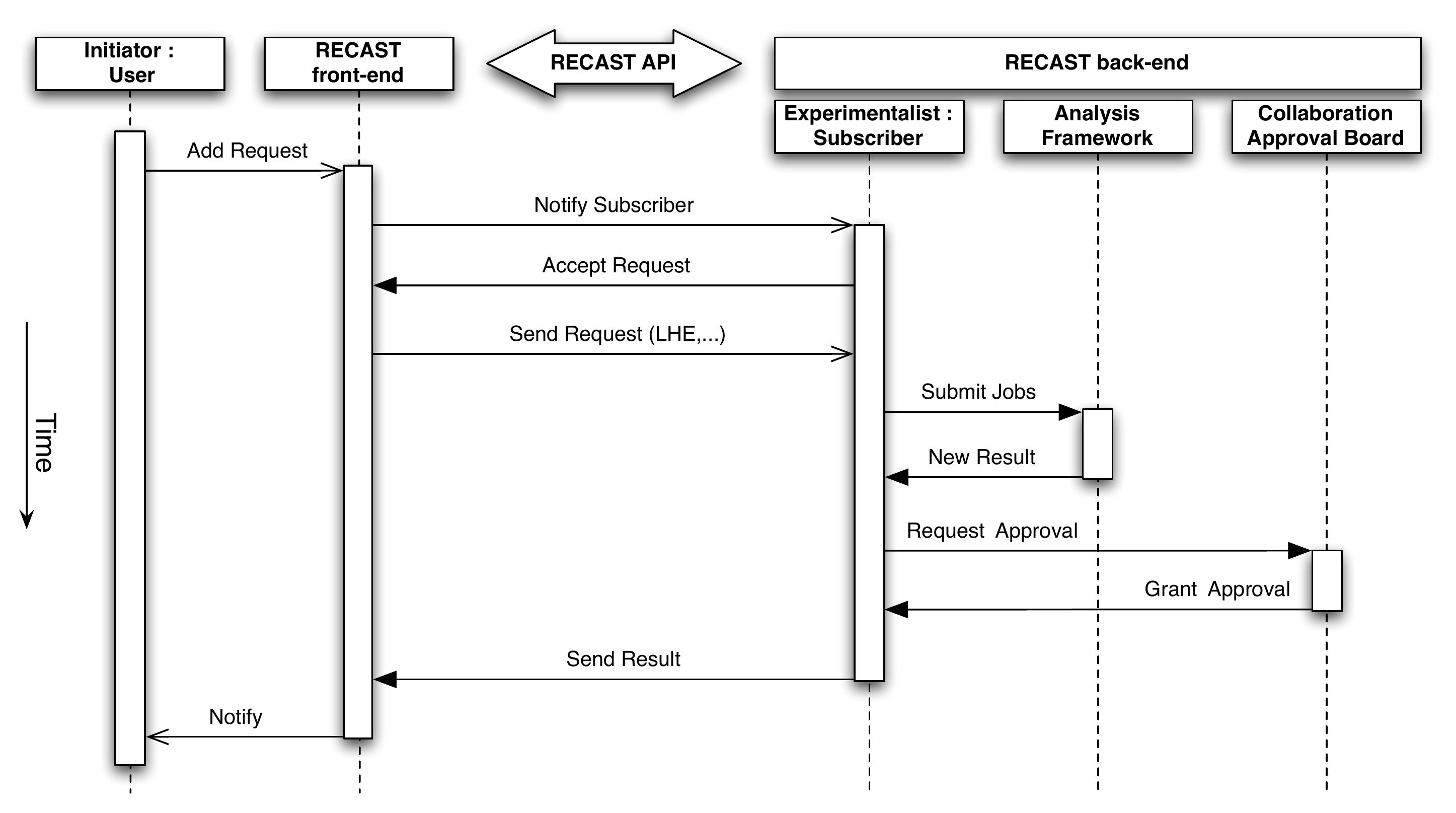}
\end{center}
\caption{A sequence diagram outlining the interactions involved in processing a request through RECAST (time flows from top to bottom).  A user initiates a request with the RECAST front-end; the front-end communicates the request to the back-end via the RECAST API; and the collaboration-specific implementation of the back-end processes the alternative signal to arrive at a new result. Note that formal approval is unnecessary for requests made internal to a collaboration.}
\label{fig:SequenceDiagram}
\end{figure}

Fig.~\ref{fig:SequenceDiagram} shows a diagram that outlines the sequence of events initiated by a new request to the RECAST front-end and ending in the notification that a new result is available.   

%
%
%
%

\subsection*{The RECAST front-end}

The front-end of the RECAST framework serves as in information broker.  It will 
\begin{itemize}
 \item present a list of existing analyses,
 \item collect requests for alternative signals to be processed by these analyses,
 \item notify subscribers to the relevant analyses when new requests have been made, and
 \item collect results of any processed requests.
\end{itemize}
We imagine that analyses will be added to the system in one of two ways.  In the first case, an experimentalist or collaboration will add the analysis to the RECAST listing and indicate that it is ``RECAST-ready" (ie. that the back-end is prepared to accept requests for alternative signals).  In the second case, an interested theorist or experimentalist will add the analysis to the listing knowing that it is relevant for their alternative signal and anticipating that requests will be accepted in the future.  The analysis listings will provide links to the arXiv and published articles and provide a subscription mechanism so that subscribers will be informed when new requests have been submitted or processed.

The submission of a new request will require several pieces of information.  The primary ingredient will be a sample of signal events in the Les Houches Accord format~\cite{Boos:2001cv,Alwall:2006yp}, which has been extremely successful in allowing the integration of different simulation software with each other.  A request is associated with a single analysis, and multiple signal samples may be required as some searches span a number of running conditions (eg. the multiple beam-energies of LEP searches). The request should also include links to references that motivate the alternative scenario and justify why the existing search may be sensitive to the alternative signal.  A reference cross-section for each sample of signal events should be provided to aid in excluding specific reference scenarios, or to relate the expected signal yield across multiple running conditions.

It is rare that an alternative signals exist in isolation, usually it corresponds to a point in the parameter space of a model for new physics.  Clearly, the front-end should anticipate scans over the parameter space of such theories.  Thus, requests will be organized into scans, and each request will also provide the list of model parameters and their values.  Conversely, multiple requests may be sent to different analyses testing the same alternative signal. Thus, the system will  provide a means to aggregate these results.  Initially, each point in the scan will require a dedicated request for each analysis, but one can imagine the front-end automating such scans by taking advantage of packages such as FeynRules~\cite{Christensen:2008py}.   Eventually, the front-end may even be able to present the results graphically in the form of 95\% confidence level contours in the corresponding parameter space.

\subsection*{The RECAST back-end}

The workhorse of the RECAST framework is the back-end associated with a particular analysis.  The back-end will
\begin{itemize}
 \item supplement the signal sample with a tuned parton-shower and hadronization, if the input is provided at the parton-level;
 \item simulate the signal events in the detector, including effects of pile-up, if necessary;
 \item reconstruct the events with the corresponding algorithms used for the original search;
 \item apply all event selection criteria used in the original search;
 \item obtain the selection efficiency and necessary distributions of the selected signal events;
 \item determine the upper- and lower-bounds on the rate of events of this type consistent with the original analysis's background estimates and observation with data; and
 \item obtain formal approval of the new results and communicate them to the front-end.
\end{itemize}
Initially, we anticipate that these steps  will all be done manually and recognize that it will require a real investment from experimentalists to answer a RECAST request.  We hope that this will evolve, and that the collaborations will develop systems so that analyses can be archived.  Below we discuss some of the considerations that have been identified through preliminary conversations with experimentalists or encountered directly in the authors' experience with the analysis of archived LEP data.


Immediately after the completion of a search, the additional effort required to process an alternative signal through the back-end is relatively modest.  The required effort grows quickly with time as analysis codes are lost and common software evolves.  Thus, the key to a successful RECAST back-end will be an archival framework.  The challenges are largely technical in nature and will be particular to each collaboration's analysis pipeline.  While it may seem daunting initially, the use of virtual machines addresses most issues that have plagued archival efforts of the past.  Since the back-end does not require access to data or large samples of simulated backgrounds, the resources necessary to archive the analysis chain are also relatively modest\footnote{Since the event selection typically targets the bulk of the signal distribution, we expect $\mathcal{O}(1000)$ events to be sufficient to obtain a reliable estimate of the efficiency for most requests.  The ATLAS computing model estimates  2-15 min (2000 kSI2K-sec) to simulate a single event with the full GEANT4 simulation on a single computing core~\cite{AtlasSimulation}.  Thus,  we anticipate a single 4-core machine to be able to process one request per day, while a moderately sized cluster to be able to process $\mathcal{O}(100)$ requests per day.  }.  Nevertheless, the successful system will require a dedicated effort and the involvement of computing professionals.

Much of the necessary analysis pipeline is already in place.  Most experiments are already capable of processing Les Houches event files through showering and hadronization generators, detector simulation, and reconstruction in an automated way using GRID tools.  In some cases, the event selection can also be incorporated into the same system.  The challenge lies in integrating the final stages of an analysis into a well-defined structure.  Because the RECAST API provides a well-defined and minimal target, it may facilitate creation of a general back-end system within each experiment.

%
%

For analyses based on a single event count, the only piece of information necessary to recast the original analysis is the selection efficiency for the alternative signal.  However, for analyses that use the distribution of event counts in multiple bins or unbinned likelihood analysis the overall signal efficiency is not sufficient information.   In those analyses where this is indeed the case, the statistical procedure used to produce confidence intervals on the signal rate will also need to be processed by the back-end.   Thus, in addition to the event selection, each analysis will need to archive the original statistical procedure or provide an alternative.  We note that in many cases, for instance likelihood ratio tests, it is natural for the final statistical test to be based on the alternative signal.

The last function of the back-end is to provide authority to the final result.  We anticipate that initially this will require someone within the collaboration to shepherd the new result through the collaboration's approval process.  This may be expedited if collaborations provide a streamlined approval process similar to CDF's ``reblessing'' procedure for updating results based on the same analysis.  We note that formal approval is unnecessary for requests made internal to a collaboration.

\subsection*{The RECAST API}

The language that enables communication between the front-end and back-end is the RECAST API (application programming interface).  It will:
\begin{itemize}
 \item define the format of requests sent to a collaboration,
 \item define the format of results obtained by a collaboration, and
 \item establish a communication protocol between the front-end and back-end.
\end{itemize}
A well-defined API is the key to enabling multiple back-ends to communicate effectively with the front-end.  It is analogous to the Les Houches interfaces which enable interoperability between the various event generators~\cite{Boos:2001cv,Skands:2003cj,Alwall:2006yp}.  Web-based APIs are also at the heart of the GRID tools that coordinate the full-chain of generation, simulation, and event reconstruction.  The initial version of the RECAST API is very light-weight, as it is restricted to the sequence diagram shown in Fig.~\ref{fig:SequenceDiagram}.  The details of this initial API are beyond the scope of this document, but will be provided together with a beta-version of the RECAST front-end paired with a toy example of a back-end~\cite{RECAST-web} .  This initial implementation is intended to be complete, but may require some modifications based on the feedback from the collaborations.

\section{Conclusions}

As we demonstrated above, existing analyses are often sensitive to a broader class of models than they were originally designed to test. When handled with care the community is willing to accept the result of an existing analysis recast into the context of an alternative signal. With the recent advancement in simulation tools it is now possible to generate and consider alternative scenarios of interest to the community on a short time-scale. Thus, it is natural to develop a framework like RECAST to fully exploit the power of existing analyses to guide the community in our search for new physics. 

The impact of RECAST depends entirely on the incorporation and integration of existing analyses into the framework. It builds on the efforts of experimental searches by extending and expanding their relevance to the community, all under the auspices of the collaborations. 
The design and considerations put forth in this work aim to initiate a community wide effort to bring such a framework to life. 
%


\acknowledgments

We would like to thank Neal Weiner, Paul Granis, Roberto Tenchini, Markus Luty, Michael Peskin, Paul de Jong, Daniel Whiteson, Tim Tait, David Tucker-Smith, Matthew Reece, Patrick Meade, and David Shih for their support and useful feedback. K.C. is supported by the US National Science Foundation grants PHY-0854724 and PHY-0955626.   I.Y. is supported by the James Arthur fellowship. 

\bibliographystyle{JHEP}
\bibliography{recast}

\begin{thebibliography}{10}
\bibitem{Caveat}
 This often consists of leading-order treatment of the hard scattering with
  tools like MadGraph, while showering and hadronization is performed with
  tools like Pythia or Herwig. When more subtle effects of QCD are important,
  the two stages must be coordinated and tools like Sherpa, MC@NLO, or POWHEG
  are often used.
\end{thebibliography}
\end{document}